\documentclass[12pt]{article}
\usepackage{times}

\usepackage{epsf}
\def\beq{\begin{equation}}
\def\eeq{\end{equation}}
\def\bea{\begin{eqnarray}}
\def\eea{\end{eqnarray}}

\def\ver{\right|}
\def\vel{\left|}
\def\ver{\right|}
\def\nnb{\nonumber}
\def\ga{\left(}
\def\dr{\right)}

\def\rar{\rightarrow}
\def\nnb{\nonumber}
\def\lla{\langle}
\def\rra{\rangle}
\def\ba{\begin{array}}
\def\ea{\end{array}}
\def\bea{\begin{eqnarray}}
\def\eea{\end{eqnarray}}

\title{Rare  $B_s \rar \gamma \,\nu \bar{\nu} $ decay with polarized photon and new physics effects}
\author{\vspace{1cm}\\
         {\bf Berin \c{S}irvanl{\i}$^a$}
         \thanks{E-mail address:
        berin@quark.fef.gazi.edu.tr} \, \, and \, \,
        {\bf G\"{u}rsevil  Turan$^b$}
        \thanks{E-mail address:
        gsevgur@metu.edu.tr}\,\,
\\
\\
{\small \sl ${}^a$Gazi University, Faculty of Arts and Science, Department of Physics,
06100, }
\\  {\small \sl  $\ $Teknikokullar, Ankara, Turkey}
\\
\\
and
\\
\\
  {\small \sl ${}^b$Physics Department, Middle East Technical University ,}
\\   {\small \sl  $\ $ 06531 Ankara, Turkey }}

\begin{document}
\maketitle

\begin{abstract}
Using the most general, model independent  effective Hamiltonian,
the rare $B_s \rar \gamma \, \nu \bar{\nu} $ decay with polarized
photon is studied. The sensitivity of the branching ratio and
photon polarization to the new Wilson coefficients are
investigated. It is shown that these physical observables  are
sensitive to the vector and tensor type interactions, which would
be useful in search of new physics.
\end{abstract}
~~~PACS number(s): 12.60.--i, 13.20.--v, 13.20.He

\newpage
\section{Introduction}
Since the first observation of the flavor--changing neutral current
induced decays $B \rar X_s \gamma$ and $B \rar K^* \gamma$ by CLEO \cite{CLEO},
the rare B-meson
decays have begun to play a more  important role in the particle physics phenomenology.
 This can be attributed to the fact that as these processes only occur at loop level in
 the weak interaction, they provide fertile
testing ground for the gauge sector of the SM. In addition, these decays
can also serve as a good probe for establishing  new physics  beyond the SM.

From experimental point of view, studying rare $B$ meson decays can provide  essential information
about  the poorly known parameters of the SM, such as the elements of the
Cabibbo--Kobayashi--Maskawa (CKM) matrix, the leptonic decay constants
etc. In this connection, the pure leptonic B decays $B\rar \ell^+\ell^-$
could be used to get such information. However, these decays are proportional to the
lepton mass, so that they are  helicity
suppressed with the factor of $m^2_{\ell}/m^2_{B}$, having the branching ratios of
$BR(B_s\rar e^+e^-,\mu^+\mu^-)\simeq 10^{-14}, 10^{-9}$ \cite{Eilam}. For $\tau$ channel,
there is no such suppression and $BR(B_s\rar \tau^+\tau^-)\simeq 10^{-7}$ \cite{Buchalla}, but
this time the difficulty arises from their experimental observation due to the low efficiency.
As for the $B \rar \nu \bar{\nu}$ decay,  it is obviously forbidden in the SM due to the helicity 
conservation if neutrino is massless.
Alternatively, the $B \rar \gamma \, \nu \bar{\nu} $
decays can be considered as potential processes to determine the above-mentioned constants,
especially the leptonic decay constant $f_{B}$, since its branching ratio (BR) quadratically
depends on this constant. Due to the additional the photon emission in
$B \rar \gamma \,\nu \bar{\nu} $,
the  helicity suppression is removed, as in $B \rar \gamma \, \ell^+ \ell^- $ decay,
so that larger BR is expected. The SM prediction for  $BR(B_s\rar  \gamma \,\nu \bar{\nu} )$
is of the order of  $\sim 10^{-8}$ and it has very clear experimental signature, i.e.,
the "missing mass" and isolated photon so that they are among the potentially
measurable decays in the near future.

In this paper, we will use the  light cone QCD sum rules method to evaluate the
hadronic matrix elements and  study the rare $B_s \rar \gamma \,\nu \bar{\nu} $ decay
in a general model independent way by taking into account the polarization effects of the photon.
$B \rar \gamma \, \nu \bar{\nu} $ decay has been investigated in the SM,
using the constituent quark model and pole models \cite{Lu} for the determination of the leptonic decay
constants $f_B$. In \cite{Aliev1}, this process was
studied also within the framework of the light cone QCD sum rules method, but there
neutrino was assumed to be massless and the Hamiltonian consisted of a single term representing the four
vector interactions of the left handed neutrinos only. However, the Super Kamiokande \cite{SK}
results indicated that neutrino had mass so that it could have right components. Therefore,
in our work, we use a most general model independent  effective Hamiltonian, which contains the scalar and
tensor type interactions as well as the vector  types (See Eq.(\ref{effH}) below).
We note that this mode has been studied  in a recent work \cite{Berin} in a similar way
to our analysis and   shown
that the spectrum is sensitive to the types of the interactions so that it is useful to
discriminate the various new physics effects.

In a radiative decay mode like ours, the final
state photon can emerge with a definite polarization and studying  the effects of polarized photon may
provide another kinematical variable, in addition to the differential and total
branching ratios for radiative decays \cite{Choud}. Although experimental measurement of this variable
would be much more difficult than that of e.g., polarization  of the final leptons in 
$B \rar \gamma \, \ell^+ \ell^- $ decay, this is still another kinematical variable for studying
radiative decays. In this work we will investigate sensitivity of  
such "photon polarization asymmetry" in $B_s \rar \gamma \, \nu \bar{\nu}$ decay 
to the new Wilson coefficients.

The paper is organized as follows: In section 2, we present the most general, model
independent form of the effective Hamiltonian and the parametrization of the
hadronic matrix elements in terms of appropriate form factors. We then calculate
the differential decay width and the  photon polarization  asymmetry for the 
$B_s \rar \gamma \,\nu \bar{\nu}$ decay when the photon is in 
positive and negative helicity states . Section 3 is devoted to the numerical analysis and
discussion of our results.

\section{Matrix element for the  $B_s  \rar \gamma \, \nu \bar{\nu} $ decay}
For the semileptonic $B_s \rar \gamma \, \nu \bar{\nu} $ decay, the basic  quark level process
is  $b \rar s \nu \bar{\nu} $, which can be written in terms of 10 model independent
four--Fermi interactions  in the following form \cite{Aliev2}-\cite{Fukae}:
\bea
\label{effH}
{\cal H}_{eff} &=& \frac{\alpha \,G_F }{4\sqrt{2}\pi}
\frac{ V_{tb}V_{ts}^\ast}{ \sin^2 \theta_W}
\Bigg\{ C_{LL}^{tot}\, \bar s \gamma_\mu (1-\gamma_5) b \,\bar \nu \gamma^\mu (1-\gamma_5)\nu +
C_{LR}^{tot} \,\bar s \gamma_\mu (1-\gamma_5) b \, \bar \nu \gamma^\mu (1+\gamma_5)\nu \nnb \\
&&+C_{RL} \,\bar s \gamma_\mu (1+\gamma_5)b \,\bar \nu \gamma^\mu (1-\gamma_5)\nu +
C_{RR} \,\bar s \gamma_\mu (1+\gamma_5)b \, \bar \nu \gamma^\mu (1+\gamma_5)\nu \nnb \\
&& +C_{LRLR} \, \bar s (1+\gamma_5) b \,\bar \nu (1+\gamma_5) \nu +
C_{RLLR} \,\bar s (1-\gamma_5) b \,\bar \nu (1+\gamma_5) \nu \nnb \\
&&+C_{LRRL} \,\bar s(1+\gamma_5) b \,\bar \nu (1-\gamma_5) \nu +
C_{RLRL} \,\bar s (1-\gamma_5) b \,\bar \nu (1-\gamma_5) \nu \nnb \\
&&+ C_T\, \bar s \sigma_{\mu\nu} b \,\bar \nu \sigma^{\mu\nu}\nu
+i C_{TE}\,\epsilon^{\mu\nu\alpha\beta} \bar s \sigma_{\mu\nu} b \,
\bar \nu \sigma_{\alpha\beta} \nu  \Bigg\}~, \nnb
\eea
where $C_X$ are the coefficients of the four--Fermi interactions with $X=LL,LR,RL,RR$
describing
vector, $X=LRLR,RLLR,LRRL,RLRL$ scalar and $X=T,TE$ tensor type interactions.
The Wilson coefficients $C_{LL}$ and $C_{LR}$ already exist in the SM in the form
$C_9^{eff} - C_{10}$ and $C_9^{eff} + C_{10}$ for the $b\rar s \ell^+\ell^-$ decay, while
for $b\rar s \nu \bar{\nu}$ transition, we have $C_9 = -C_{10}=X(x_t)$  \cite{inami} with
\bea
X(x_t) & = & \frac{x_t}{8}\Bigg[ \frac{x_t+2}{x_t-1}+\frac{3 (x_t-2)}{(x_t-1)^2}
\ln(x_t)\Bigg]+\frac{\alpha_s}{4 \pi}\, X_1(x_t) \, .
\eea
Here,  $x_t=m^2_t/m^2_W$ and $X_1(x_t)$, which gives about $3\%$ contribution to $C_9$, can be found in
\cite{Buchalla}. Now, writing
\bea
C_{LL}^{tot} &=& 2 X(x_t) + C_{LL}~, \nnb \\
C_{LR}^{tot} &=&  C_{LR}~, \nnb
\eea
we see that $C_{LL}^{tot}$ contains the contributions from the SM and also from the new
physics. We note that the Wilson coefficients $C_{LRLR}$, $C_{RLLR}$, $C_{LRRL}$ and $C_{RLRL}$
in Eq.(\ref{effH}) describe the scalar type interactions, which do not contribute for 
the $B_s \rar \gamma \, \nu \bar{\nu} $ processes.

The next step is, starting with the Hamiltonian in Eq.(\ref{effH}), to calculate the
$B_s \rar \gamma \, \nu \bar{\nu} $ decay at hadronic level. We already note that the pure
leptonic decay
$B_s \rar \nu \bar{\nu} $ is forbidden due to helicity conservation. However, when
a photon is emitted from initial b or s-quark lines, this pure leptonic process change
into  the corresponding radiative one and no helicity suppression exists anymore.
The contributions coming from the release of a free photon from any internal charged line
will be suppressed by a factor of $m^2_b/m^2_W$, therefore they can be neglected. Thus, the matrix
elements necessary to calculate $B_s \rar \gamma \,\nu \bar{\nu} $ decay are as follows
\cite{Eilam2}-\cite{Aliev4}:

\bea
\label{mel1}
\lla \gamma(k) \vel \bar s \gamma_\mu
(1 \mp \gamma_5) b \ver B(p_B) \rra &=&
\frac{e}{m_B^2} \Big\{
\epsilon_{\mu\nu\lambda\sigma} \varepsilon^{\ast\nu} q^\lambda
k^\sigma g(q^2) \nnb \\
&&\pm i\,
\Big[ \varepsilon^{\ast}_{\mu} (k q) -
(\varepsilon^\ast q) k_\mu \Big] f(q^2) \Big\}~,\\ \nnb \\
\label{mel2}
\lla \gamma(k) \vel \bar s \sigma_{\mu\nu} b \ver B(p_B) \rra &=&
\frac{e}{m_B^2}
\epsilon_{\mu\nu\lambda\sigma} \Big[
G \varepsilon^{\ast\lambda} k^\sigma +
H \varepsilon^{\ast\lambda} q^\sigma +
N (\varepsilon^\ast q) q^\lambda k^\sigma \Big]~,\\
\lla \gamma(k) \vel \bar q
(1 \mp \gamma_5) b \ver B(p_B) \rra &=& 0 \label{ff3}~,
\eea
where $\varepsilon_\mu^\ast$ and $k_\mu$ are the four vector polarization
and four momentum of the photon, respectively, $q$ is the momentum transfer
and $p_B$ is the momentum of the $B$ meson. Due to Eq.(\ref{ff3}), the scalar type interactions
do not contribute to the $B_s \rar \gamma \,\nu \bar{\nu} $ decay, as we have noted before.  
Using eqs.(\ref{mel1}-\ref{ff3}), the matrix element for $B_s \rar \gamma \,\nu \bar{\nu} $
decay can be calculated as follows:
\bea
\label{sd}
{\cal M} &=& \frac{\alpha G_F}{4 \sqrt{2} \, \pi} \frac{V_{tb} V_{ts}^*}{\sin^2 \theta_W}
\frac{e}{m_B^2} \,\Bigg\{
\bar \nu \gamma^\mu (1-\gamma_5) \nu \, \Big[
A_1 \epsilon_{\mu \nu \alpha \beta}
\varepsilon^{\ast\nu} q^\alpha k^\beta +
i \, A_2 \Big( \varepsilon_\mu^\ast (k q) -
(\varepsilon^\ast q ) k_\mu \Big) \Big] \nnb \\
&+& \bar \nu \gamma^\mu (1+\gamma_5) \nu \, \Big[
B_1 \epsilon_{\mu \nu \alpha \beta}
\varepsilon^{\ast\nu} q^\alpha k^\beta
+ i \, B_2 \Big( \varepsilon_\mu^\ast (k q) -
(\varepsilon^\ast q ) k_\mu \Big) \Big] \nnb \\
&+& i \, \epsilon_{\mu \nu \alpha \beta}
\bar \nu \sigma^{\mu\nu}\nu \, \Big[ G \varepsilon^{\ast\alpha} k^\beta
+ H \varepsilon^{\ast\alpha} q^\beta +
N (\varepsilon^\ast q) q^\alpha k^\beta \Big] \\
&+& i \,\bar \nu \sigma_{\mu\nu}\nu \, \Big[
G_1 (\varepsilon^{\ast\mu} k^\nu - \varepsilon^{\ast\nu} k^\mu) +
H_1 (\varepsilon^{\ast\mu} q^\nu - \varepsilon^{\ast\nu} q^\mu) +
N_1 (\varepsilon^\ast q) (q^\mu k^\nu - q^\nu k^\mu) \Big] \Bigg\}~,\nnb
\eea
where
\bea
A_1 &=& ( C_{LL}^{tot} + C_{RL} ) g ~~~,~~~ \nnb \\
A_2 &=& ( C_{LL}^{tot} - C_{RL} ) f ~, \nnb \\
B_1 &=& ( C_{LR}^{tot} + C_{RR} ) g ~, \nnb \\
B_2 &=& ( C_{LR}^{tot} - C_{RR} ) f ~, \nnb \\
G &=& 4 C_T g_1 ~, \nnb \\
N &=& - 4 C_T \frac{1}{q^2} (f_1+g_1) ~, \\
H &=& N (qk) ~, \nnb \\
G_1 &=& - 8 C_{TE} g_1 ~, \nnb \\
N_1 &=& 8 C_{TE} \frac{1}{q^2} (f_1+g_1) ~, \nnb \\
H_1 &=& N_1(qk) \nnb
\eea

The next task is the calculation of the  differential decay
rate of $B_s \rar \gamma \, \nu \bar{\nu} $ decay  as a function of
dimensionless parameter $x=2 E_{\gamma}/m_B$, where $E_{\gamma}$
is the  photon energy. In such a   radiative decay, the final
state photon can emerge with a definite polarization and there follows the question of how 
sensitive the branching ratio is to the new Wilson coefficients when the photon is in the 
positive or negative helicity states.  To find an answer to this question, we  evaluate 
$\frac{d\Gamma (\varepsilon^\ast=\varepsilon_1)}{dx}$ and 
$\frac{d\Gamma (\varepsilon^\ast=\varepsilon_2)}{dx}$ for $B_s \rar \gamma \, \nu \bar{\nu} $ decay, 
in  the c.m. frame of $\nu \bar{\nu}$, in which four-momenta and
polarization vectors ,$\varepsilon_1$ and $\varepsilon_2$, are as follows:
\bea
P_B & = & (E_B,0,0,E_k) \,\,\, , \,\,\, k = (E_k,0,0,E_k) \,\,\, , \,\,\,
p_1=(p,0,p \sqrt{1-z^2},-p z) \nnb \\ p_2 & = & (p,0,-p \sqrt{1-z^2},p z)\,\,\, , \,\,\,
\varepsilon_1 = (0,1,i,0)/\sqrt{2}\,\,\, , \,\,\,\varepsilon_2 = (0,1,-i,0)/\sqrt{2}
\label{mom}
\eea
where $E_B=m_B (2-x)/2 \sqrt{1-x}$, $E_k=m_B x/2 \sqrt{1-x}$, and $p=m_B \sqrt{1-x}/2$.
In Eq.(\ref{mom}) $z=\cos \theta$, where $\theta$ is the angle between the momentum of the B-meson and
that of $\nu$ in the c.m. frame of $\nu \bar{\nu}$. Using the above forms, obtain
\bea
\label{bela} \frac{d\Gamma (\varepsilon^\ast=\varepsilon_i)}{dx}& =& \,\vel \frac{\alpha G_F}{4
\sqrt{2} \, \pi} \frac{V_{tb} V_{ts}^*}{\sin^2 \theta_W} \ver^2 \,
\frac{\alpha}{\ga 2 \, \pi \dr^3}\,\frac{\pi}{4}\,m_B \, \Delta(\varepsilon_i)
\eea
where
\bea 
\label{deltapm}
\Delta(\varepsilon_i) & = &\frac{1}{3} x \,
\Bigg\{  (1-x)\Bigg[ x^2\,m_B^2\Big( \vel A_1 \ver^2 + \vel A_2 \ver^2 +
\vel B_1 \ver^2 + \vel B_2 \ver^2\pm 2 \,\mbox{\rm Re} [A_1 A_2^\ast+B_1 B_2^\ast]\Big)
\nnb \\
&-&4 \Big( (-1+x) (\vel H_1 \ver^2 + \vel H \ver^2)+x(\pm
\mbox{\rm Im} [G_1 H^\ast-G H_1^\ast]- \mbox{\rm Re} [G_1 H_1^\ast+G H^\ast]) \Big)   \Bigg]
\nnb \\ & + & 2 x^2\Big( \vel G_1 \ver^2 + \vel G \ver^2 \pm 2 \, \mbox{\rm Im} [-G_1 G^\ast] \Big)\Bigg\}
\eea
with $+(-)$ is for $i=1(2)$.

The effects of polarized photon can be also studied through a variable "photon polarization asymmetry",
\cite{Choud}:
\bea
H(x)=\frac{\frac{d\Gamma (\varepsilon^\ast=\varepsilon_1)}{dx}-
\frac{d\Gamma (\varepsilon^\ast=\varepsilon_2)}{dx}}{\frac{d\Gamma (\varepsilon^\ast=\varepsilon_1)}{dx}+
\frac{d\Gamma (\varepsilon^\ast=\varepsilon_2)}{dx}}
& = & \frac{\Delta(\varepsilon_1)-\Delta(\varepsilon_2)}{\Delta_0} \, ,
\eea
where 
\bea
\label{Hx}
\Delta(\varepsilon_1)-\Delta(\varepsilon_2) = 
\frac{4}{3}  x^2 \Bigg\{(1-x)
\Big( m^2_B  x \mbox{\rm Re} [A_1 A_2^\ast+B_1 B_2^\ast] +
2 \mbox{\rm Im} [G H_1^\ast-G_1 H^\ast]\Big)-2 x \mbox{\rm Im} [G_1 G^\ast]\Bigg\}
\eea
and
\bea 
\Delta_0 & = & x^3 \,
\Bigg\{\frac{2}{3} m_B^2 (1-x)\,\Big( \vel A_1 \ver^2 + \vel A_2 \ver^2 +
\vel B_1 \ver^2 + \vel B_2 \ver^2\Big)+\frac{4}{3} \,\Big( \vel G_1 \ver^2 + \vel G \ver^2 \Big)\nnb \\
&+&4 \frac{(1-x)}{x^2}\Big( (1-x) (\vel H_1 \ver^2 + \vel H \ver^2)+x
\mbox{\rm Re} [G_1 H_1^\ast+G H^\ast] \Big)\nnb \\
 &-&  \frac{1}{3} m_B^2 (1-x)    \Big( 2  \mbox{\rm Re} [G N^\ast+G_1 N_1^\ast]  +
 m^2_B (1-x)  \Big( \vel N_1 \ver^2 + \vel N \ver^2 \Big)    \Big)  \Bigg\}
\eea

\section{Numerical analysis and discussion}
In this section we will present our numerical analysis about the branching ratio (BR) and the photon
polarization asymmetry $H$ for $B_s \rar \gamma \,\nu \bar{\nu} $ decay.
It follows from Eqs. (\ref{bela}) and (\ref{Hx}) that in order to make such numerical predictions,
we first of all need the explicit forms of the form factors $g,~f,~g_1$ and $f_1$.
They are calculated in framework of light--cone $QCD$ sum rules
in \cite{Eilam2,Aliev3}, in terms of two parameters $F(0)$ and $m_F$ as
\begin{eqnarray}
F(q^2) & = & \frac{F(0)}{\left(1-\frac{(1-x)m^2_B}{m^2_F}\right)^2}
\end{eqnarray}
where the values  $F(0)$ and $m_F$ for the $B_s \rightarrow \gamma$ transition
are listed in Table 1.
\begin{table}[h]
\renewcommand{\arraystretch}{1.5}
\addtolength{\arraycolsep}{3pt}
$$
\begin{array}{|l|cc|}
\hline
& F(0) & m_F \\ \hline
g &
\phantom{-}1 \, GeV & 5.6 \, GeV \\
f &
\phantom{-}0.8 \, GeV & 6.5 \, GeV \\
g_1 &
 \phantom{-}3.74 \, GeV^2 & 6.4 \, GeV \\
f_1&
  \phantom{-}0.68 \, GeV^2 & 5.5 \, GeV \\
\hline
\end{array}
$$
\caption{$B$ meson decay form factors in the light-cone QCD sum rule.}
\renewcommand{\arraystretch}{1}
\addtolength{\arraycolsep}{-3pt}
\end{table}

The values of the other input parameters which have been used in the present work are:
$m_{B_s}=5.28~GeV$, $\tau (B_s)=1.61 \times 10^{-12}~s$,
$\vel V_{tb} V_{ts}^\ast \ver = 0.045$, $\alpha^{-1}=137$,
$G_F=1.17\times 10^{-5}~GeV^{-2}$, $C_9=1.618$. Furthermore we assume in this work that
all new Wilson coefficients are real and vary in the region $-4\leq C_X\leq 4$.
We note that such a  choice for the range of the new Wilson coefficients follows from
the experimental bounds on the branching ratios of the $B_s\rar K^{\ast}\mu^+\mu^-$ and
$B_s\rar \mu^+\mu^-$ decays \cite{affolder}.

For reference, we first present our SM prediction, $BR(B_s \rar \gamma \nu \bar{\nu})=9.54\times 10^{-8}$,
which is in a good agreement with the result of ref. \cite{Aliev1}.

\begin{figure}[htb]
\vskip 0truein \centering \epsfxsize=3.8in
\leavevmode\epsffile{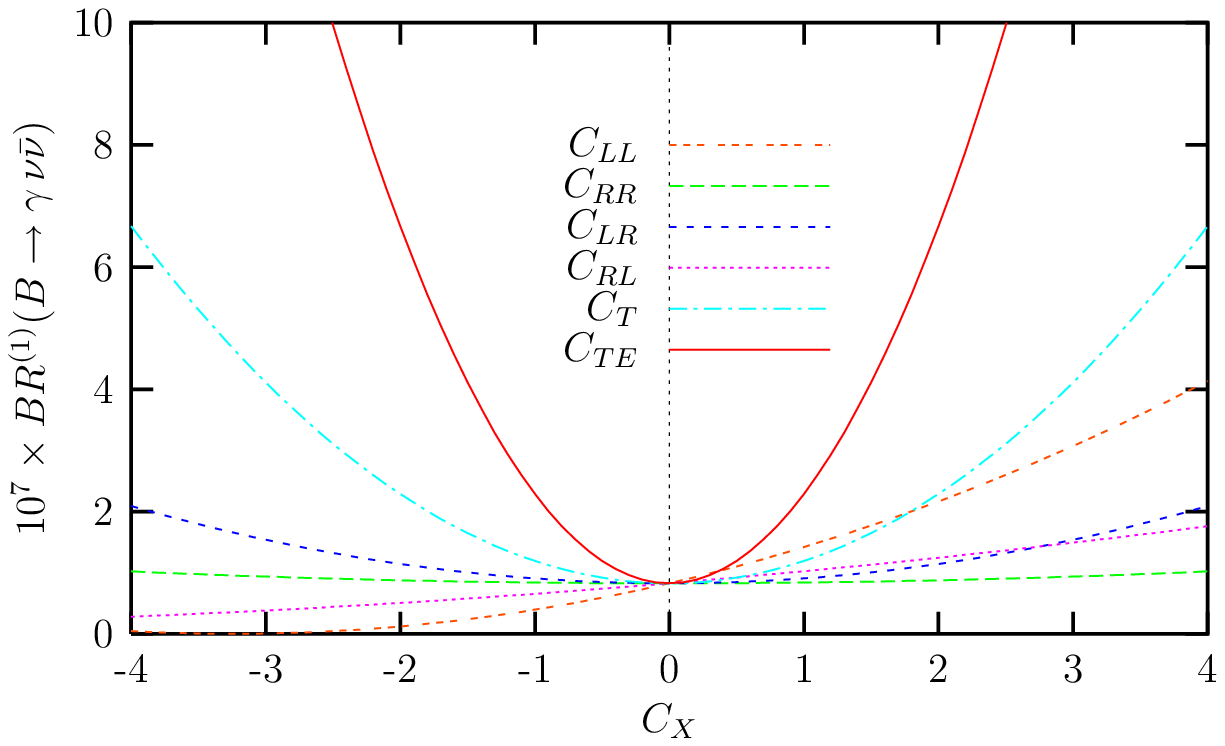} \vskip 0truein
\caption[]{The dependence of the integrated branching ratio $BR^{(1)}$ on the new Wilson coefficients
for the $B_s \rar  \gamma \nu \bar{\nu}$ decay. The superscript $(1)$ corresponds to the photon in
positive helicity state.}\label{f1}
\end{figure}

\begin{figure}[htb]
\vskip 0truein \centering \epsfxsize=3.8in
\leavevmode\epsffile{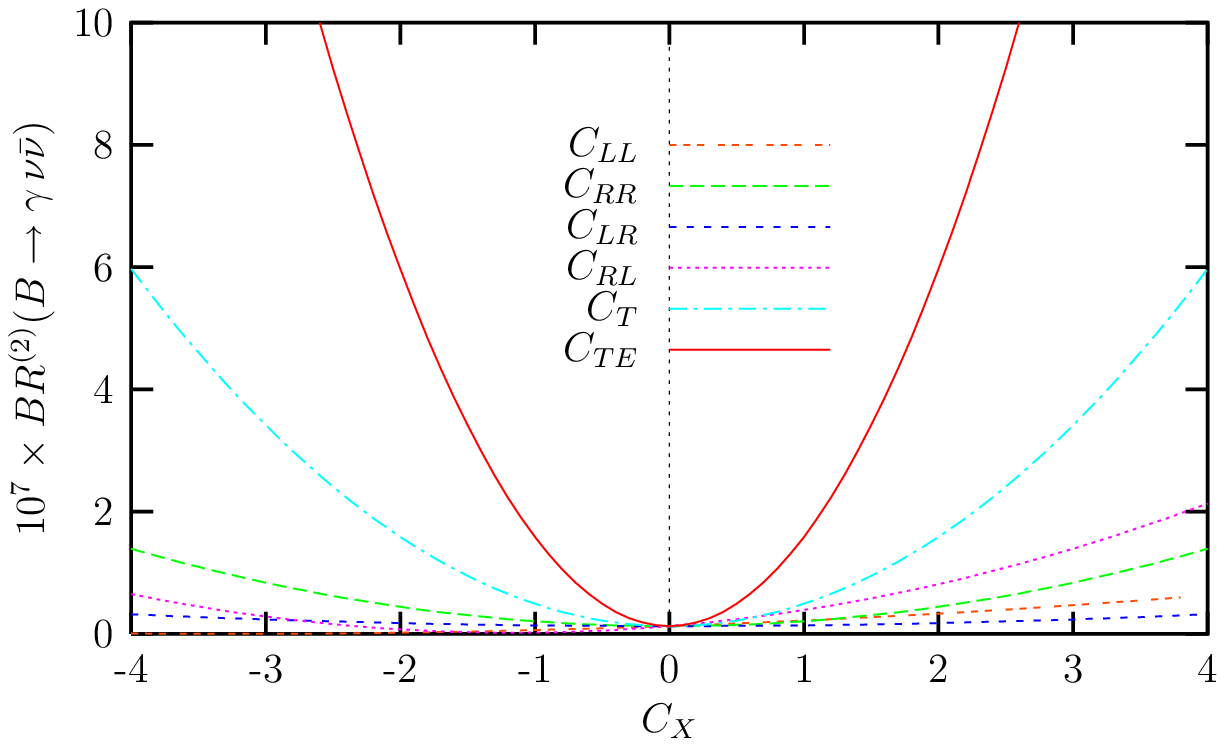} \vskip 0truein
\caption[]{The same as Fig.(\ref{f1}), but for the photon in negative helicity state.}\label{f2}
\end{figure}

In Figs. (\ref{f1}) and (\ref{f2}), we present the dependence of the 
$BR^{(1)}$ and $BR^{(2)}$  for $B_s \rar \gamma \nu \bar{\nu}$ decay
on the new Wilson coefficients, $C_{LL}$,$C_{RR}$,$C_{LR}$,$C_{RL}$,$C_{T}$ and $C_{TE}$, where the 
superscripts $(1)$ and $(2)$ correspond to the positive and negative helicity states of photon, respectively.
We observe  from these figures that among the  new interactions  considered in the effective Hamiltonian
the branching ratio   in both cases is most sensitive to the tensor
type of interactions. Indeed, e.g., for $C_{T}(C_{TE})=\pm 2$, $BR^{(1)}$ is larger about $3(8)$ times
compared to that of the SM prediction. For $BR^{(2)}$, we get even larger enhancement, which is 
about $12(45)$ times compared to the SM. In addition, the dependence of the branching ratio in both cases
on $C_{T}$ and $C_{TE}$ is symmetrical with respect to the zero point. From Figs.(\ref{f1}) and (\ref{f2}),
we also see that branching ratio $BR^{(1)}$
is  sensitive to the vector interactions with coefficients $C_{LL}$ and $C_{LR}$, while $BR^{(2)}$
is more sensitive to the coefficients $C_{RR}$ and $C_{RL}$. 

\begin{figure}[htb]
\vskip 0truein \centering \epsfxsize=3.8in
\leavevmode\epsffile{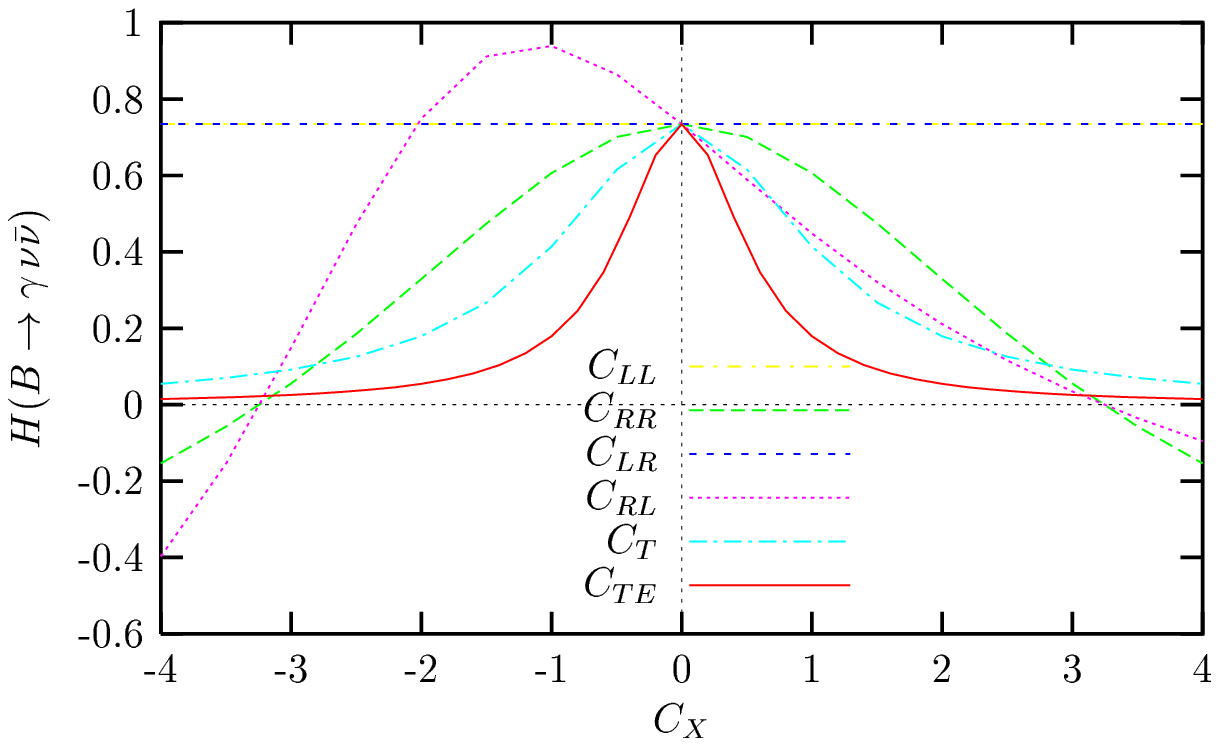} \vskip 0truein
\caption[]{The dependence of the integrated photon polarization asymmety of
the $B_s \rar  \gamma \nu \bar{\nu}$  decay on the new Wilson coefficients.}
 \label{f3}
\end{figure}

\begin{figure}[htb]
\vskip 0truein \centering \epsfxsize=3.8in
\leavevmode\epsffile{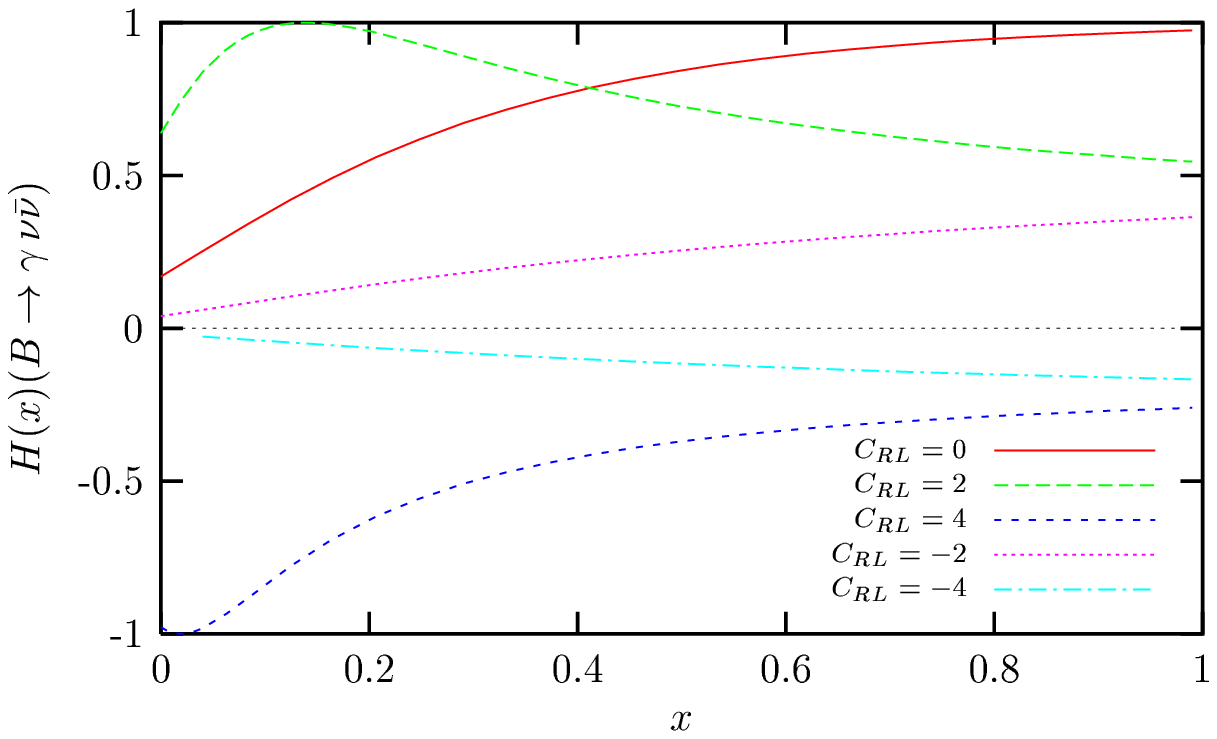} \vskip 0truein
\caption[]{The dependence of the differential  photon polarization asymmety of
the $B_s \rar  \gamma \nu \bar{\nu}$  decay on $x$ for different values of
$C_{RL}$.} \label{f4}
\end{figure}

\begin{figure}[htb]
\vskip 0truein \centering \epsfxsize=3.8in
\leavevmode\epsffile{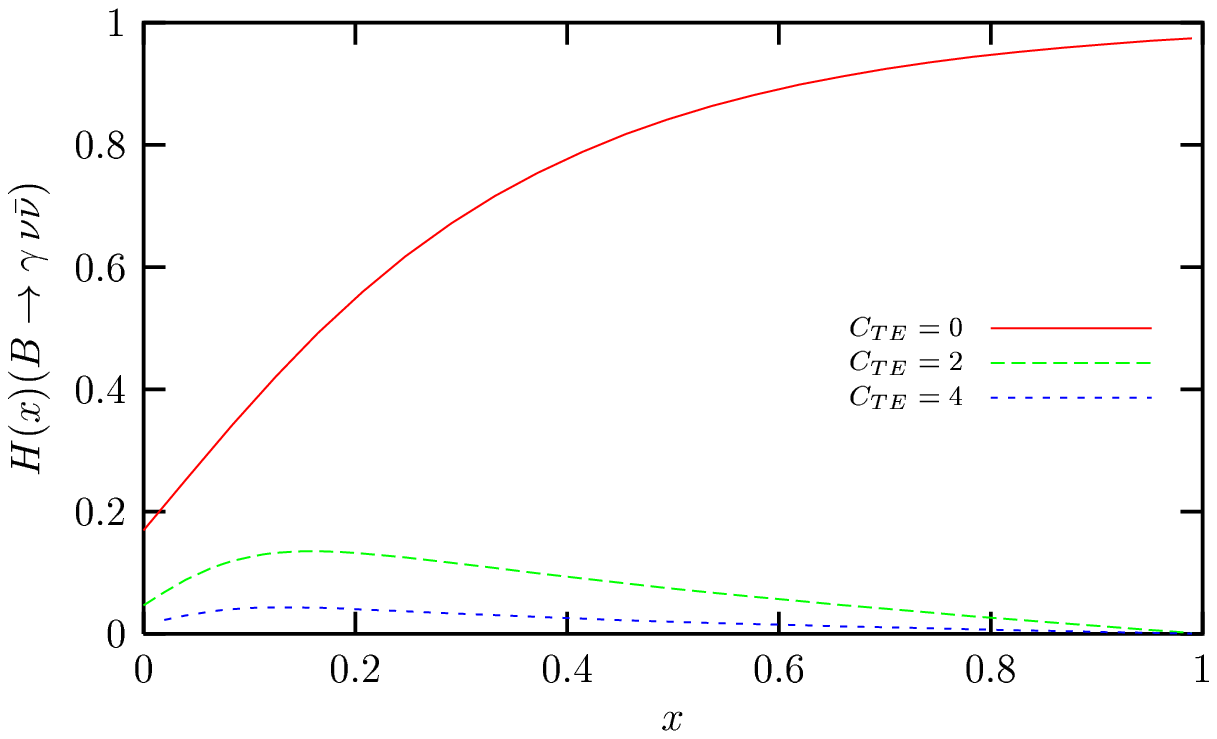} \vskip 0truein
\caption[]{The dependence of the differential  photon polarization asymmety of
the $B_s \rar  \gamma \nu \bar{\nu}$  decay on $x$ for different values of
$C_{TE}$.} \label{f5}
\end{figure}

In Fig.(\ref{f3}), we present the dependence of the integrated photon polarization
asymmetry $H$  for $B_s \rar \gamma \nu \bar{\nu}$ decay on the new Wilson coefficients,
$C_{LL}$,$C_{RR}$,$C_{LR}$,$C_{RL}$,$C_{T}$ and $C_{TE}$. We see that spectrum of $H$
is perfectly symmetrical with respect to the zero point for all the new Wilson coefficients,
except the $C_{RL}$. The coefficient $C_{RL}$, when it is between $-2$ and $0$,
is also the only one  which gives the constructive contribution to the SM prediction
of $H$, which we find $H(B_s \rar \gamma \nu \bar{\nu})=0.73$. This behavior is
clear also from  Fig.(\ref{f4}), in which
we plot the differential photon polarization asymmetry $H(x)$ for the same decay
as a function of dimensionless  variable $x=2E_{\gamma}/m_B$ for the different values
of the vector  interaction with coefficients $C_{RL}$. Finally, we give in Fig.(\ref{f5}),
$H(x)$ as a function of $x$ for  the different values
of the tensor  interaction with coefficients $C_{TE}$. From these two last figures,
we can conclude that performing measurement of $H$ at different photon energies
can give information about the signs of the new Wilson coefficients, as well
as their magnitudes.

In conclusion, we have studied the branching ratio of the
$B_s \rar \gamma \nu \bar{\nu}$ decay  when photon has positive and negative helicities,  
also another physically measurable quantity, namely the photon polarization asymetry of the
same decay, using a general, model independent effective Hamiltonian.
It follows that  these physical observables are very sensitive to the existence of new
Wilson coefficients so that their experimental measurements can give valuable information 
about new physics.

\vspace{1cm}

We would like to thank T. M. Aliev and M. Savc{\i} for useful discussions.

\newpage

\end{document}